\documentclass{emulateapj}
\usepackage{apjfonts}
\usepackage{mathrsfs}

\usepackage{amsmath}

\usepackage{graphics}

\newcommand{\bd}{\begin{displaymath}}
\newcommand{\ed}{\end{displaymath}}

\newcommand{\beq}{\begin{equation}}
\newcommand{\eeq}{\end{equation}}
\newcommand{\bea}{\begin{eqnarray}}
\newcommand{\eea}{\end{eqnarray}}

\newcommand{\cs}{c_{\rm s}}

\newcommand{\rmd}{{\rm d}}
\newcommand{\ncd}{\kern -1pt\cdot\kern -2pt }
\newcommand{\ecd}{\kern -2pt\cdot\kern -1pt }
\newcommand{\Pm}{\mathrm{P_m}}

\slugcomment{accepted by ApJ}

\shorttitle{The large scale magnetic fields of thin accretion disks}

\begin{document}

\title{The large scale magnetic fields of thin accretion disks }

\author{Xinwu Cao\altaffilmark{1} and Hendrik C. Spruit\altaffilmark{2}}

\affil{$^{1}$ Key Laboratory for Research in Galaxies and Cosmology,
Shanghai Astronomical Observatory, Chinese Academy of Sciences, 80
Nandan Road, Shanghai, 200030, China; cxw@shao.ac.cn}

\affil{$^{2}$ Max Planck Institute for Astrophysics,
Karl-Schwarzschild-Str. 1, 85748, Garching, Germany;
henk@mpa-garching.mpg.de}

\begin{abstract}
Large scale magnetic field threading an accretion disk is a key
ingredient in the jet formation model. The most attractive scenario
for the origin of such a large scale field is the advection of the
field by the gas in the accretion disk from the interstellar medium
or a companion star. However, it is realized that outward diffusion
of the accreted field is fast compared to the inward accretion
velocity in a geometrically thin accretion disk if the value of the
Prandtl number $\Pm$ is around unity. In this work, we revisit this
problem considering the angular momentum of the disk is removed
predominantly by the magnetically driven outflows. The radial
velocity of the disk is significantly increased due to the presence
of the outflows. Using a simplified model for the vertical disk
structure, we find that even moderately weak fields can cause
sufficient angular momentum loss via a magnetic wind to balance
outward diffusion. There are two equilibrium points, one at low
field strengths corresponding to a plasma-beta at the midplane of
order several hundred, and one for strong accreted fields,
$\beta\sim 1$. We surmise that the first is relevant for the
accretion of weak, possibly external, fields through the outer parts
of the disk, while the latter one could explain the tendency,
observed in full 3D numerical simulations, of strong flux bundles at
the centers of disk to stay confined in spite of strong MRI
turbulence surrounding them.
\end{abstract}



\section{Introduction}

{\tt \obeylines}

Jets/outflows are observed in different types of the sources, such
as, active galactic nuclei (AGNs), X-ray binaries, and young stellar
objects, which are probably driven from the accretion disk through
the magnetic field lines threading the disk \citep*[see reviews
in][]{1996epbs.conf..249S,2000prpl.conf..759K,2007prpl.conf..277P,
2010LNP...794..233S}. The large scale magnetic field co-rotates with
the gases in the disk, and the jets/outflows are powered by the
gravitation energy released by accretion of the gases through the
ordered field threading the disk. A large scale magnetic
field, of uniform polarity threading the inner parts of the disk is probably
a key ingredient in this jet formation model.

The origin of such a field
is not well understood, however, since the net magnetic flux threading a
disk cannot be produced or changed by internal processes in the disk (a
consequence of the solenoidal nature of the magnetic field, see e.g.\ Spruit 2010).
A net magnetic flux in the inner disk must therefore either be inherited
from initial conditions, or somehow be accreted from a larger distance;
ultimately for example from the interstellar medium or a companion star
(cf. Bisnovatyi-Kogan \& Ruzmaikin 1974, 1976).

One could imagine a steady state
in which the inward advection of the field lines is balanced by the outward
movement of field lines due to magnetic diffusion.
In conventional isotropic idealizations of a turbulent plasma, it is expected
(e.g.\ Parker 1979) that $\nu\sim\eta\sim lv_{\rm t}$, in which $l$ is the
largest eddy size, and $v_{\rm t}$ is turnover velocity, i.e.\ $\Pm\sim
1$. Whether this is actually the case in MRI turbulence has been investigated
using numerical simulations \citep*[e.g.,][]{2003A&A...411..321Y,
2009A&A...504..309L,2009A&A...507...19F,2009ApJ...697.1901G}. The
results all suggest that the effective magnetic Prandtl number is around
unity. Fromang et al.\ (2009),  for example, measure $\Pm\approx 2$.
For such Prandtl numbers a vertical field (perpendicular to the disk plane)
can indeed be dragged efficiently by an advection dominated accretion flow
\citep{2011ApJ...737...94C}, which is hot and geometrically thick
\citep{1994ApJ...428L..13N,1995ApJ...452..710N}.

The advection of the field in a geometrically thin ($H/r\ll1$), turbulent
accretion disk is inefficient, however, because the radial
component of the magnetic field diffuses much faster across the disk, on a time
scale $\sim H^2/\eta$. As a result, a magnetic field with inclination
$B_r/B_z\sim 1$ actually diffuses outward on a time scale of order
$H/r$ shorter than a purely vertical field (van Ballegooijen 1989).
An inclined field is a necessary consequence of effective accretion
of the field, however, since the accumulation of field lines in the
inner disk exerts a pressure that causes the field above the disk
to spread outward. While such an inclined configuration is
favorable for launching a flow \citep*{1982MNRAS.199..883B,
1994A&A...287...80C}, it raises the problem how it can be accreted
effectively against the action of magnetic diffusion.

Several alternatives were suggested to resolve the difficulty of
field advection in thin accretion disks. \citet{2005ApJ...629..960S}
suggested that a weak large-scale magnetic field threads the disk in
the form of localized patches in which the field is strong enough to
cause efficient angular momentum through a magnetic wind. General
relativistic magnetohydrodynamic (GRMHD) simulations of an accretion
torus embedded in a large-scale magnetic field showed that a central
magnetic flux bundle, once formed from a suitable initial condition,
can survive in spite of MRI turbulence present in the disk
surrounding it \citep{2009ApJ...707..428B}. The calculations by
\citet{2012MNRAS.424.2097G} and Guilet \& Ogilvie (2012b) show that
the accretion velocity of the gas in the region away from the
midplane of the disk can be larger than that at the midplane of the
disk, which may partially solve the problem of too efficient
diffusion of the field in thin accretion disk.

The outward diffusion of the field could be balanced by accretion if a process can be found that increases the accretion velocity by a factor $\sim r/H$ relative to the rate due to the turbulence alone. We explore here the conditions under this can be achieved by a magnetic wind generated by the weak magnetic field that is to be accreted.

\section{Model}
\subsection{Model assumptions}

Apart from the accreted field, the disk model we use is a standard
$\alpha$--disk model, i.e. with a viscosity $\nu$ parametrized as
$\nu=\alpha c_{\rm s} H$, where $c_{\rm s}$ is the (isothermal)
sound speed, $H$ is the scale height of the disk, and $\alpha\sim
0.01-0.1$ (the range of values measured in MRI simulations). The
field to be accreted is assumed to be sufficiently weak that it does
not suppress magnetororational instability. The disk thus contains
an MRI-generated field as well as a weaker field of uniform polarity
threading the disk.

Since the accretion velocity that is to be achieved exceeds the
viscous rate, we simplify the analysis by assuming the accretion
flow to be dominated by the angular momentum loss from the accreted
field, ignoring the viscous contribution from the MRI turbulence.
The resulting accretion rate then has to exceed outward diffusion
due to the MRI turbulence. The turbulence is assumed to produce an
effective magnetic diffusivity corresponding to a magnetic Prandtl
number $\Pm \sim 1$.

The model needs a prescription for the angular momentum loss
produced by the accreting weak magnetic field. For this we use the
Weber-Davis model for a magnetically driven wind, in the `cold'
approximation (in which the gas pressure force is neglected). It
leads to a simple description in terms of the field strength and
mass loss rate (Mestel 2012, Spruit 1996). The Weber-Davis model
strictly applies only to the `split monopole' configuration, in
which the poloidal field is purely radial. Its properties are found
to be a rather good approximation for more general poloidal field
shapes as well (Anderson et al. 2005), which makes it adequate for
the present purpose.

The angular momentum loss in this model can be characterized by a
single dimensionless constant, a `mass load parameter'$\mu$ (c.f.\  Michel 1969, Mestel, 2012):
\beq \mu =\chi \, 4\pi\Omega r_0/ B_{\rm p},\eeq
where $B_{\rm p}=(B_z^2+B_r^2)^{1/2}$ is the poloidal component of the accreted field at radius $r_0$ on the surface of the disk, and $\Omega$ the rotation rate at $r_0$ (cylindrical coordinates $r,\phi,z$). $\chi$ is a constant along the flow, it is a measure of the `mass flux per field line':
\beq \chi=\rho v_{\rm p}/ B_{\rm p},\eeq
where $\rho$, $v_{\rm p}$, $B_{\rm p}$ are the mass density, poloidal velocity and poloidal field strength. It is related to the mass flux per unit surface area from the disk $\dot m_{\rm w}$ by
\beq
\dot m_{\rm w}=B_z \chi.
\eeq
The asymptotic velocity of the wind decreases monotonically with increasing $\mu$, and  for $\mu=1$ equals the rotation velocity at $r_0$. The angular momentum loss increases monotonically with $\mu$.

It turns out that the conditions for effective accretion of the field can be satisfied when $\mu\ga \alpha\, \Pm\, r/H$. This is described in the following sections, using a specific model for the disk structure.

\subsection{The magnetic field of the disk}

The angular momentum equation for a steady accretion disk with
outflows is
\begin{equation}
{\frac {d}{dr}}(2\pi r \Sigma v_r r^2\Omega)={\frac {d}{dr}}(2\pi r
\nu \Sigma r^2{\frac {d\Omega}{dr}})-2\pi rT_{\rm m},
\label{angular_1}
\end{equation}
where  $\Sigma$ is the total surface mass density of the disk (counting both sides),  $T_{\rm m}$ the total torque per unit of surface area of the disk (counting both sides) due to the magnetically driven outflows. Integrating Eq.\ (\ref{angular_1}), yields
\begin{equation}
2\pi r\Sigma v_r r^2\Omega=2\pi r\nu\Sigma r^2{\frac
{d\Omega}{dr}}-2\pi f_{\rm m}(r)+C, \label{angular_2}
\end{equation}
where the value of the integral constant $C$ can be determined with a
boundary condition on the accreting object, and
\begin{equation}
rT_{\rm m}={\frac {df_{\rm m}(r)}{dr}}. \label{f_m}
\end{equation}
The magnetic torque in unit surface disk area exerted by the
outflows is
\begin{equation}
T_{\rm m}=2\dot{m}_{\rm w} r_{\rm A}^2\Omega, \label{t_m_1}
\end{equation}
where $\dot{m}_{\rm w}$ is the mass loss rate in the outflow from
a unit surface area of the disk (single sided), $r_{\rm A}$ is the (cylindrical)
Alfv{\'e}n radius of the outflow, $\Omega(r)$ is the angular velocity of the
disk, and $r$ is the radius of the field line foot point
at the disk surface. The dimensionless mass load parameter $\mu$ of the outflow is
\begin{equation}
\mu={\frac {4\pi\rho_{\rm w}v_{\rm w}\Omega r}{B_{\rm p}^2}}={\frac
{4\pi\Omega r}{B_{\rm p} B_z}}\dot{m}_{\rm w}, \label{mu_1}
\end{equation}
where $\rho_{\rm w}v_{\rm w}=\dot m_{\rm w} B_{\rm p}/B_z$ is the mass flux
parallel to the field line. In the cold Weber-Davis model the Alfv{\'e}n radius is
\begin{equation}
r_{\rm A}=r\left[{\frac {3}{2}}(1+\mu^{-2/3})\right]^{1/2}.
\label{r_a_1}
\end{equation}
[For more detailed discussion of this model see Mestel (2012) or
\citet{1996epbs.conf..249S}].  MHD simulations of axisymmetric
magnetically driven flows have shown that relations like (\ref{r_a_1}) and
(\ref {t_m_2}) below are fair approximations for more general
configurations than the Weber-Davis `split monopole'  \citep*[cf.\ Fig.\  7
in][]{2005ApJ...630..945A}. Substituting Eqs.\ (\ref{r_a_1}) and
(\ref{mu_1}) into (\ref{t_m_1}), we find
\begin{equation}
T_{\rm m}={\frac {3}{4\pi}r B_{\rm p}^2\mu(1+\mu^{-2/3})}.
\label{t_m_2}
\end{equation}
The radial velocity of an accretion flow in which the angular momentum is
 removed predominantly by the outflows can be estimated from Eq.\  (\ref{angular_1}):
\begin{equation}
v_r\sim -{\frac {T_{\rm m}}{\Sigma}}\left[{\frac {d}{dr}}(r^2\Omega)
\right]^{-1}\simeq-{\frac {2T_{\rm m}}{\Sigma r\Omega}},
\label{v_r_1}
\end{equation}
where we have assumed that the rotation is approximately Keplerian, $\Omega\approx\Omega_{\rm K}$. [This is sufficient for the following estimates, but has to be made more precise when considering the wind launching conditions (section \ref{massflux}).] The mass accretion rate of the accretion disk is
\begin{equation}
\dot{M}=-2\pi r\Sigma v_r\simeq {\frac {4\pi T_{\rm m}}{\Omega}},
\label{mdot_d_1}
\end{equation}
where Eq.\ (\ref{v_r_1}) is used. We can compare the mass loss
rate in the outflows with the accretion rate through the disk. Using
 (\ref{mu_1}), (\ref{t_m_2}) and (\ref{mdot_d_1}):
\begin{equation}
{\frac {d\ln\dot{M}}{d\ln r}}={\frac {4\pi r^2\dot{m}_{\rm
w}}{\dot{M}}}={\frac {1}{3}}(1+\mu^{-2/3})^{-1}.\label{mdot_d_2}
\end{equation}
This shows that for large mass loading parameters, $\mu\ga 1$, a fraction $\sim 1/3$ of the accretion rate can be lost in the wind, if the wind is sustained across the disk over a distance of order $r$. For $\mu\la 1$ the mass loss in the wind is not significant compared with the accretion rate.
Substituting (\ref{t_m_2}) into (\ref{v_r_1}), we obtain
\begin{equation}
v_r=-{\frac {6 c_{\rm s}g(\mu)}{\Omega H\beta_{\rm p}}}c_{\rm s},
\label{v_r_2}
\end{equation}
where $g=\mu(1+\mu^{-2/3})$,
$c_{\rm s}$ is the sound speed of the gas in the midplane of the disk, and
\begin{equation}
\beta_{\rm p}=p_{\rm c}/{\frac {B_{\rm p}^2}{8\pi}}
\end{equation}
is a measure of the poloidal field strength at the surface relative to $p_{\rm c}$,
the gas pressure at the midplane of the disk.
The magnetic field is advected inwards on a timescale
\begin{equation}
\tau_{\rm adv}\sim {\frac {r}{|v_r|}}=
{r\over c_{\rm s}}{\Omega H\beta_{\rm p}\over 6 c_{\rm s}g(\mu)}.
\label{adv}
\end{equation}
In the limit $H/r\ll 1$, the radial component of the field dominates
the outward diffusion timescale $\tau_{\rm diff}$ of the magnetic
field  (van Ballegooijen 1989, Lubow et al.\ 1994). Let
\beq\kappa_0=B_z/B_{r,{\rm s}}\label{kappa_0}\eeq be the
inclination, with respect to the horizontal, of the accreted field
at the disk surface (s). The diffusion time scale is then of order
\begin{equation}
\tau_{\rm dif}\sim {\frac {r H\kappa_0}{\eta }},\label{tau_dif_1}
\end{equation}
where $\eta$ is the magnetic diffusivity. [This dependence holds as long
as $1/\kappa_0 > H/r$; in the opposite case of a nearly vertical field $B_r/B_0 <H/r$, the
diffusion time is of order $r^2/\eta$].
The magnetic Prandtl number is defined as $\Pm=\eta/\nu$, where $\nu$
is the turbulent viscosity. With the conventional $\alpha$-parametrization,
$\nu=\alpha c_{\rm s}H$ we can then write Eq.\ (\ref{tau_dif_1}) as
\begin{equation}
\tau_{\rm dif}\sim {r\kappa_0\over c_{\rm s}} \frac {1}{\alpha \Pm}. \label{tau_dif_2}
\end{equation}
For a steady state, the advection of the field in the disk has to
balance the diffusion, i.e. $\tau_{\rm adv}=\tau_{\rm dif}$. With
(\ref{adv}), (\ref{tau_dif_2}) this yields a condition on the mass flux parameter $\mu$:
\begin{equation} \mu(1+\mu^{-2/3})=
{\alpha\Omega H\beta_{\rm p}\Pm\over 6 c_{\rm s}\kappa_0},
\label{mu_adv_1g}
\end{equation}
which reduces to
\begin{equation}
\mu(1+\mu^{-2/3})= {\alpha\beta_{\rm p}\Pm\over 6\kappa_0}
\label{mu_adv_1}
\end{equation}
in weak field approximation, $c_{\rm s}=\Omega H$. If a significant
mass flux, $\mu\sim 1$ can be launched, and assuming $\alpha=0.01$,
P$_{\rm m}=1$, and a field inclination of 30$^\circ$ to the
vertical, this shows that a weak field with $\beta\sim 10^3$ can
still be accreted through the angular momentum loss of the wind
associated with it. In the following sections we investigate this
with specific models for the structure of the disk.

\subsection{The mass flux in the wind}
\label{massflux}
The mass loss rate in the outflow is governed by the gas density at the position of the
sonic point (in the following labeled with index $_{\rm s}$), where the flow speed equals the sound speed (more accurately, the slow mode
cusp speed). The mass flux, parallel to the field, is thus approximately $\rho c_{\rm s,s}$, and per unit of surface area parallel to the disk surface, the mass flux is
\begin{equation}
\dot{m}_{\rm w}\sim {\frac {B_z}{B_{\rm p}}}\rho c_{\rm s,s}~, \label{mdotw}
\end{equation}
where $B_{\rm p}=(B_z^2+B_r^2)^{1/2}$ as before.
As mentioned, we consider only the wind associated with the accreted field.
The MRI turbulence also produces a field, but being a local process, we assume
that it does not produce a large scale field that would be significant for
wind-driven angular momentum loss.

The mass flux depends sensitively on the surface temperature $T_{\rm s}$. Assume
a radiative disk, i.e. the vertical energy transport through the disk is by radiation. In
the diffusion approximation, the surface temperature is then related to the
temperature $T_{\rm c}$ of the disk at the midplane of the disk by
\begin{equation}
{\frac {4\sigma T_{\rm c}^4}{3\tau}}=\sigma T_{\rm s}^4,
\end{equation}
where $\tau$ is the optical depth of the disk in the vertical
direction. In terms of the sound speeds $c_{\rm s,s}$, $c_{\rm s,c}$
at the surface and the midplane,
\begin{equation}
c_{\rm s,s}=\left({\frac{4}{3\tau}}\right)^{1/8}c_{\rm s,c}~. \label{theta_surf_1}
\end{equation}
Along a field line that corotates with its footpoint in the disk the outflow is
governed by an effective potential $\Psi_{\rm eff}$,
\begin{equation}
\Psi_{\rm eff}(r,z)={\frac {GM}{(r^2+z^2)^{1/2}}}-{\frac
{1}{2}}\Omega^2 r^2, \label{Psi_eff_1}
\end{equation}
where $\Omega$ is the angular velocity of the footpoint
at the disk surface. In this expression and in the following, the (small) difference
between the rotation rate $\Omega$ of the field line and the Keplerian value
$\Omega_{\rm K}$  has to be included consistently, since it has a strong
effect on the launching conditions for the wind,
through its effect on $\Psi_{\rm eff}$ (Ogilvie and Livio, 2001).
This difference, due to the magnetic stress exerted at the surfaces
follows from the radial equation of motion,
\begin{equation}
r\Omega_{\rm K}^2-r\Omega^2={\frac {B_{r,{\rm s}}B_{z}}{2\pi\Sigma
}}={\frac {B_{z}^2}{2\pi\Sigma \kappa_0}}, \label{omega_1}
\end{equation}
where $\Sigma$ the total (two-sided) surface density
$\Sigma\approx 2\rho_{\rm c}H$. This yields
\begin{equation}
\Omega=\Omega_{\rm K}\left[1-{\frac{2r}{\beta_z\kappa_0 H }}{\frac
{c_{\rm s}^2}{r^2\Omega_{\rm K}^2}}\right]^{1/2},\label{omega_2g}
\end{equation}
which can be approximated as
\begin{equation}
\Omega=\Omega_{\rm K}\left[1-{H\over r}{\frac{2}{\beta_z \kappa_0
}}\right]^{1/2},\label{omega_2}
\end{equation}
in weak field case, where
$\beta_z=8\pi p_{\rm c}/B_z^2=\beta_{\rm
p}(1+\kappa_0^2)/\kappa_0^2$ is the plasma-beta of the accreted
field evaluated at the midplane ($_{\rm c}$) of the disk.

Next, we make the model more specific and simplify it a bit by approximating the
sound speed in the launching region as constant with height. In addition we take the
midplane temperature as representative for the interior of the disk. This is sufficient
for the evaluation of quantities like the rotation rate correction in (\ref{omega_2}).

The location of the sonic point is close to the maximum of the effective potential.
The mass flux is determined by the density at the sonic point; a fair approximation
for this is to treat the subsonic region as if it were in hydrostatic equilibrium. This yields the
following estimate for the mass loss rate in an isothermal outflow (per unit of disk
surface area),
\begin{equation}
\dot{m}_{\rm w}\sim
 {\frac{\kappa_0}{(1+\kappa_0^2)^{1/2}}}\rho_{\rm 0} c_{\rm
s,s}\exp\left[-(\Psi_{\rm eff,s}-\Psi_{\rm eff,0})/c_{\rm s,s}^2
\right], \label{mdot_w_2}
\end{equation}
where $\rho_{\rm 0}$ is the density of the gas in the base of the outflow
(still to be specified), and $\Psi_{\rm eff,s}$ and $\Psi_{\rm eff,0}$ are the
effective potential at the sonic point and the footpoint
at the disk surface, respectively.
The factor involving $\kappa_0$ is equal to the ratio $B_z/B_{r,{\rm
s}}$ in Eq.\ (\ref{kappa_0}).

The vertical structure of an isothermal disk can be calculated with
the vertical momentum equation,
\begin{equation}
c_{\rm s}^2{\frac {\rmd\rho(z)}{\rmd z}}=-\rho(z)\Omega_{\rm K}^2 z-{\frac
{B_r(z)}{4\pi}}{\frac {\rmd B_r(z)}{\rmd z}}.\label{vertical_1}
\end{equation}
In principle, the field line shape is computable by solving the
radial and vertical momentum equations with suitable boundary
conditions (for a detailed discussion see Cao \& Spruit 2002,
hereafter CS02). For the isothermal case, an
approximate analytical expression is proposed for the shape of the
field lines in the flow:
\begin{equation}
r-r_{\rm i}={\frac {H}{\kappa_0\eta_{\rm i}^2}}(1-\eta_{\rm
i}^2+\eta_{\rm i}^2z^2H^{-2})^{1/2}-{\frac {H}{\kappa_0\eta_{\rm
i}^2}}(1-\eta_{\rm i}^2)^{1/2}, \label{b_shape_1}
\end{equation}
where $r_{\rm i}$ is the radius of the field line footpoint at the
midplane of the disk, and $\eta_{\rm i}=\tanh(1)$
(see CS02). This expression reproduces the basic
features of the Kippenhahn-Schl\"{u}ter model for a sheet of gas
suspended against gravity by a magnetic field
\citep{1957ZA.....43...36K}, for weak as well as strong field
cases. As done in CS02, we use a fitting
formula to calculate the scale height of the disk in the rest of
this work,
\begin{equation}
{\frac {H}{r}}={\frac {1}{2}}\left ({\frac {4c_{\rm
s,c}^2}{r^2\Omega_{\rm K}^2}}+{f}^2\right)^{1/2}-{\frac {1}{2}}f,
\label{h_1}
\end{equation}
where
\begin{equation}
f={\frac {1}{2(1-{\rm e}^{-1/2})\kappa_0}}{\frac
{B_z^2}{4\pi\rho\, rH\Omega_{\rm K}^2\kappa_0}}. \label{f_1}
\end{equation}
Solving for $H$:
\begin{equation}
H={c_{\rm s,c}\over\Omega_{\rm K}}\left[1-{\frac {1}{(1-{\rm e}^{-1/2})
\beta_{\rm p}(1+\kappa_0^2)}}\right]^{1/2}, \label{h_2}
\end{equation}
which requires
\begin{equation}
\beta_{\rm p}>\beta_{\rm p,min}={\frac {1}{(1-{\rm
e}^{-1/2})(1+\kappa_0^2)}}, \label{beta_condi_1}
\end{equation}
or
\begin{equation}
\beta_z>\beta_{z,\rm min}={\frac {1}{(1-{\rm e}^{-1/2})\kappa_0^2}}.
\label{betaz_condi_1}
\end{equation}
The square bracket in (\ref{h_2}) gives the magnetic correction to
the standard relation between disk thickness and the sound speed at
the midplane. It reduces to unity for a weak field, or when the
radial component is small compared with the vertical component. For
a typical value of $\kappa_0\sim 1$, we find $\beta_{\rm p}$ must be
$\ga 1.3$. This means that the magnetic pressure can be larger than
the gas pressure in the disk only if the inclination of the field
line at the disk surface $\kappa_0$ is larger than unity.

Equation\ (\ref{h_1}) should be a good approximation for the present
investigation, especially in the thin outer region of the disk which
is probably the most critical region for the accretion of a net
magnetic flux. Substituting (\ref{mdot_w_2}) into (\ref{mu_1}), the
mass flow parameter is
\begin{equation}
\mu={\frac {4\pi\Omega
r\rho_0 c_{\rm s,s}}{B_{\rm p}^2}}\exp\left[-(\Psi_{\rm
eff,s}-\Psi_{\rm eff,0})/c_{\rm s,s}^2 \right]. \label{mu_launch_1}
\end{equation}

To estimate the density $\rho_0$ at the base of the flow, we note
that Eq.\ (\ref{mu_launch_1}) is applicable only at heights in the
atmosphere where the field is strong enough, relative to the plasma,
to enforce corotation so the effective potential $\Psi$ is relevant
for the launching process of the wind. Inside the disk, where $\beta
> 1$,  this is not the case. As base of the flow, where the pressure
is $p_0$, we assume the height where the plasma-beta is of order
unity, i.e. \beq \beta_{\rm s}\equiv 8\pi p_0/B_{\rm p}^2 \approx
1,\eeq which determines $p_0$ if $\beta_{\rm s}$ is given. For most
of calculations reported below, $\beta_{\rm s}=1$ is used, some with
a lower value 0.1. The mass flux is then related to the poloidal
field strength at the disk surface by \beq \rho_0 c_{\rm
s,s}={B_{\rm p}^2\over 8\pi c_{\rm s,s}}.\eeq Using
(\ref{theta_surf_1}) Eq.\ (\ref{mu_launch_1}) can then be written as
\begin{equation}
\mu ={\Omega r\over 2 c_{\rm s,c}}(3\tau/4)^{1/8}
\exp\left[-(\Psi_{\rm eff,s}-\Psi_{\rm eff,0})/c_{\rm s,s}^2 \right]
\label{mu_launch_2}.
\end{equation}
The factor in front of the exponential tends to be a large number,
the exponential itself a small one. To evaluate the effective potential
as a function of height, field line inclination, and the (slightly non-Keplerian)
rotation rate $\Omega$, the model of CS02 is used. It is also used for the
optical depth connecting the surface temperature to the disk midplane
temperature.

\section{Results}

The disk is compressed in the vertical direction by the curved
magnetic field line, which sets an upper limit on the magnetic field
strength. We plot the corresponding minimal $\beta_z$ as a function
of the field inclination $\kappa_0$ at the disk surface in Fig.\
\ref{beta_min}. The curved field line also exerts a radial force on
the disk against the gravity of the central object, which provides
an additional constraint on the field strength (Fig.\ \ref{beta_min}
show the result for different values of the disk temperature). The
angular velocity $\Omega$ deviates from the Keplerian value due to
the radial magnetic force.  Fig.\ \ref{beta_omega} shows the
dimensionless angular velocity $\Omega/\Omega_{\rm K}$  of the disk
as a function of $\kappa_0$ and $\Theta=c_{\rm
s,c}^2/(r^2\Omega_{\rm K}^2)$.

For given values of the disk parameters, i.e., the viscosity
$\alpha$, the temperature parameter $\Theta$ in the disk, and the
optical depth $\tau$, the dependence of mass loading $\mu$ on
$\beta_z$ can be calculated with Eqs.\ (\ref{mu_adv_1g}) and
(\ref{mu_launch_2}) respectively. These relations are plotted in
Figs.\ \ref{beta_mu_th0p1}-\ref{beta_mu_th0p01} for disk-outflow
systems with different values of the parameters. In all
calculations, $\Pm=1$ is adopted. It is found that two branches of
solutions usually exist for most cases. We plot the solutions of the
disk-outflow systems in Figs.\ \ref{kappa_al0p1} and
\ref{kappa_al1p0}. The solutions with different values of
$\beta_{\rm s}$ ($\beta_{\rm s}=8\pi p_0/B_{\rm p}^2$) at the base
of the outflow for the isothermal disk are compared in Figs.\
\ref{kappa_al0p1i} and \ref{kappa_al1p0i}. The two branches of
solutions correspond to low mass ($\mu\ll 1$) loaded outflows with
strong field strength (low-$\beta$), or high mass loaded outflows
with relative weak field strength. The optical depth needed to
connect surface and internal temperature has been computed by
including a Rosseland mean opacity and electron scattering; it is
shown in Fig.\ \ref{disk_tau}.

For comparison, we have repeated the calculations for a uniformly
isothermal disk, that is,  the temperatures of disk and wind are
assumed to be the same (Figs.\ \ref{kappa_al0p1i} and
\ref{kappa_al1p0i}). Though this is not very realistic, it gives an
impression of the sensitivity of the results to the model
assumptions made.

\vskip 1cm
 \figurenum{1}
\centerline{\includegraphics[angle=0,width=7.5cm]{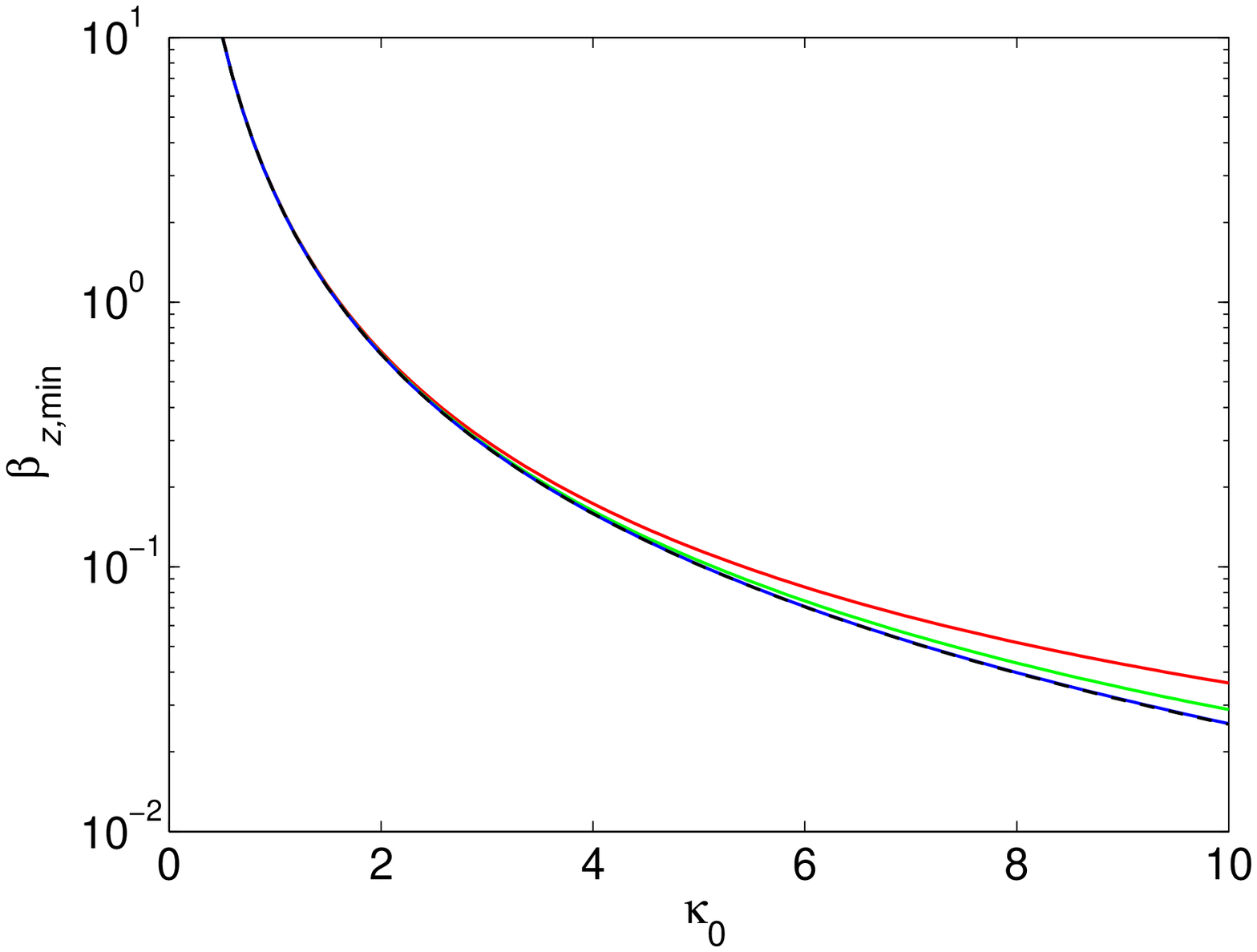}}
\figcaption{Constraints on the poloidal magnetic field strength in
terms of $\beta_z$ (the plasma-beta of the accreted field measured
at the midplane of the disk, see text). Black line: $\beta_{z,\rm
min}$ as constrained by the vertical pressure exerted by the curved
field line (Eq.\ \ref{betaz_condi_1}). Colors: the minimal values of
$\beta_{\rm p}$ constrained by the radial magnetic force (see Eq.\
\ref{omega_2g}) for different disk temperatures, $\Theta=0.01$
(red), $2.5\times 10^{-4}$ (green), and $10^{-4}$ (blue).
\label{beta_min}  }\centerline{}
 \vskip 1cm
 \figurenum{2}
\centerline{\includegraphics[angle=0,width=7.5cm]{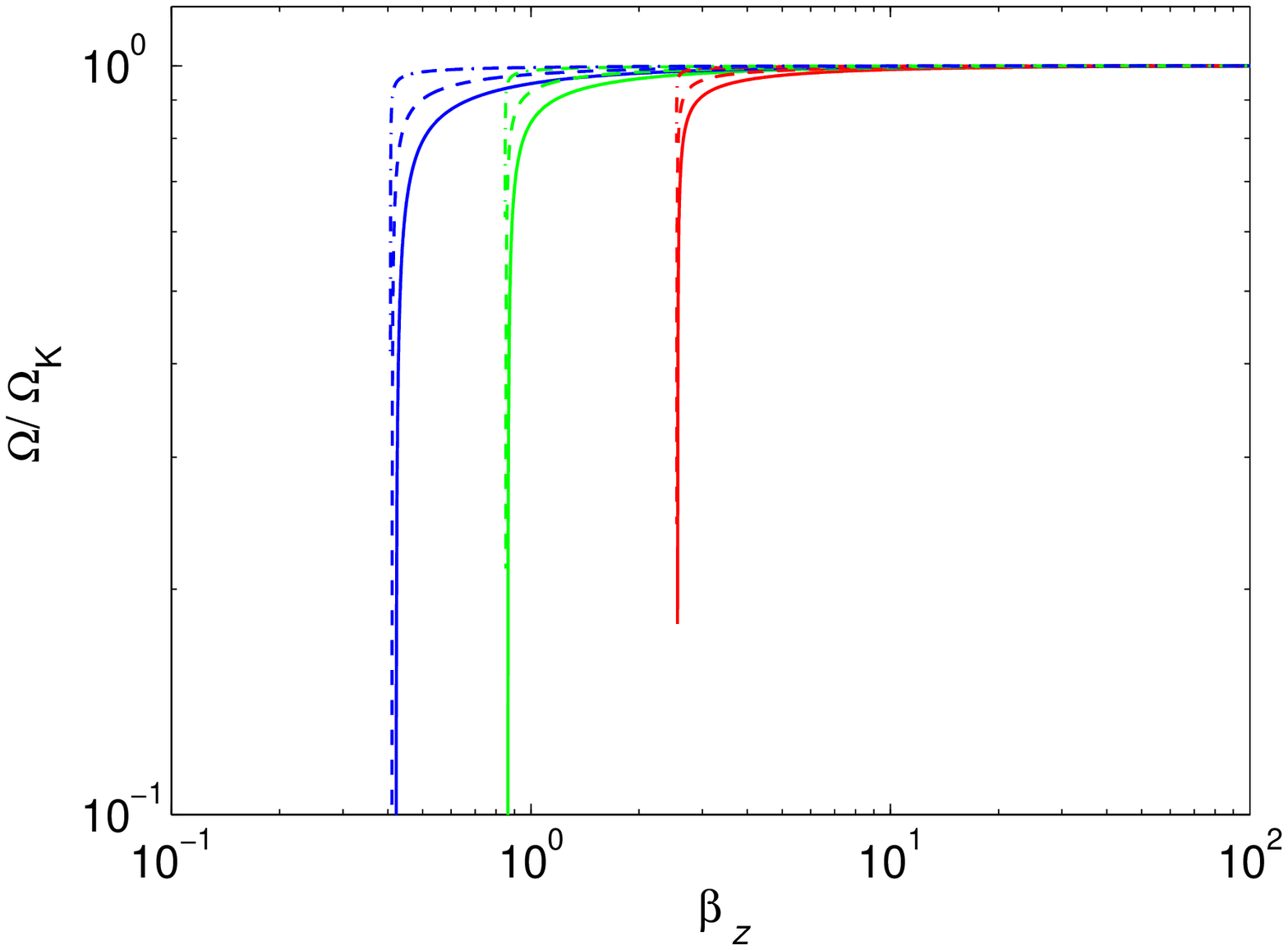}}
\figcaption{Deviation of the angular velocity of the disk ${\Omega}$
relative to Keplerian due to magnetic stress, as a function of
$\beta_z$ (Eq.\ \ref{omega_2g}). Field line inclination $\kappa_0=1$
(red), $\sqrt{3}$ (green), and $2.5$ (blue). Disk temperature
parameter $\Theta=0.01$ (solid), $2.5\times 10^{-4}$ (dashed), and
$10^{-4}$ (dash-dotted).\label{beta_omega} }\centerline{}

\subsection{Discussion}

The magnetic field is dragged inwards by the accretion disk, an
outflow is launched by this field, and the radial velocity of the
accretion disk is determined by the rate of angular momentum carried
away by the outflows. This loss rate in the outflows can be
estimated by exploring the launching process of the outflow, which
depends sensitively on the field strength/configuration and the
density/temperature of the gas at the disk surface. We have used the
cold Weber-Davis model for determining the mass load parameter $\mu$
of the outflow as a function of strength of the accreted field and
parameters of the disk structure:  its optical depth (a radiative
disk is assumed), a temperature-parameter $\Theta$, the
$\alpha$-viscosity, and the magnetic Prandtl number $\Pm$ of the
assumed MRI turbulence (cf.\ Eq.\ \ref{mu_adv_1g}). Balancing the
resulting accretion velocity with the outward diffusion by magnetic
turbulence determines the conditions for existence of a stationary
disk-outflow system. Figs.\ \ref{beta_mu_th0p1}-\ref{beta_mu_th0p01}
illustrate the properties of the solutions. It is found that two
solution branches exist for all cases. The lower branch corresponds
to high field strength and low $\mu$, i.e., low mass loss rate, the
upper one corresponds to low field strength and high $\mu$, i.e.
high mass loss rate (see Figs.\ \ref{kappa_al0p1} and
\ref{kappa_al1p0}). There is an upper limit on the field inclination
$\kappa_0$ at the disk surface, which increases with disk
temperature.  Overcoming the deeper effective potential barrier
associated with a larger inclination requires a higher internal
energy of the gas. The maximum inclination $\kappa_0$ thus decreases
with increasing optical depth $\tau$ of the disk, as the surface
temperature decreases with increasing $\tau$, (keeping other disk
parameters fixed, see Eq.\ \ref{theta_surf_1}).

We have also explored the sensitivity to model assumptions somewhat
with solutions for a more drastically simplified case, where the
temperature is assumed uniform throughout, i.e.  the temperature of
the outflow is the same as the disk temperature. This is shown in
Figs.\ \ref{kappa_al0p1i} and \ref{kappa_al1p0i}. These also show
the effect  of assuming a lower density of the gas at the base of
the outflow ($\beta_{\rm s}=0.1$). The results are qualitatively
similar to those with $\beta_{\rm s}=1$.

\vskip 1cm \figurenum{3}
\centerline{\includegraphics[angle=0,width=7.5cm]{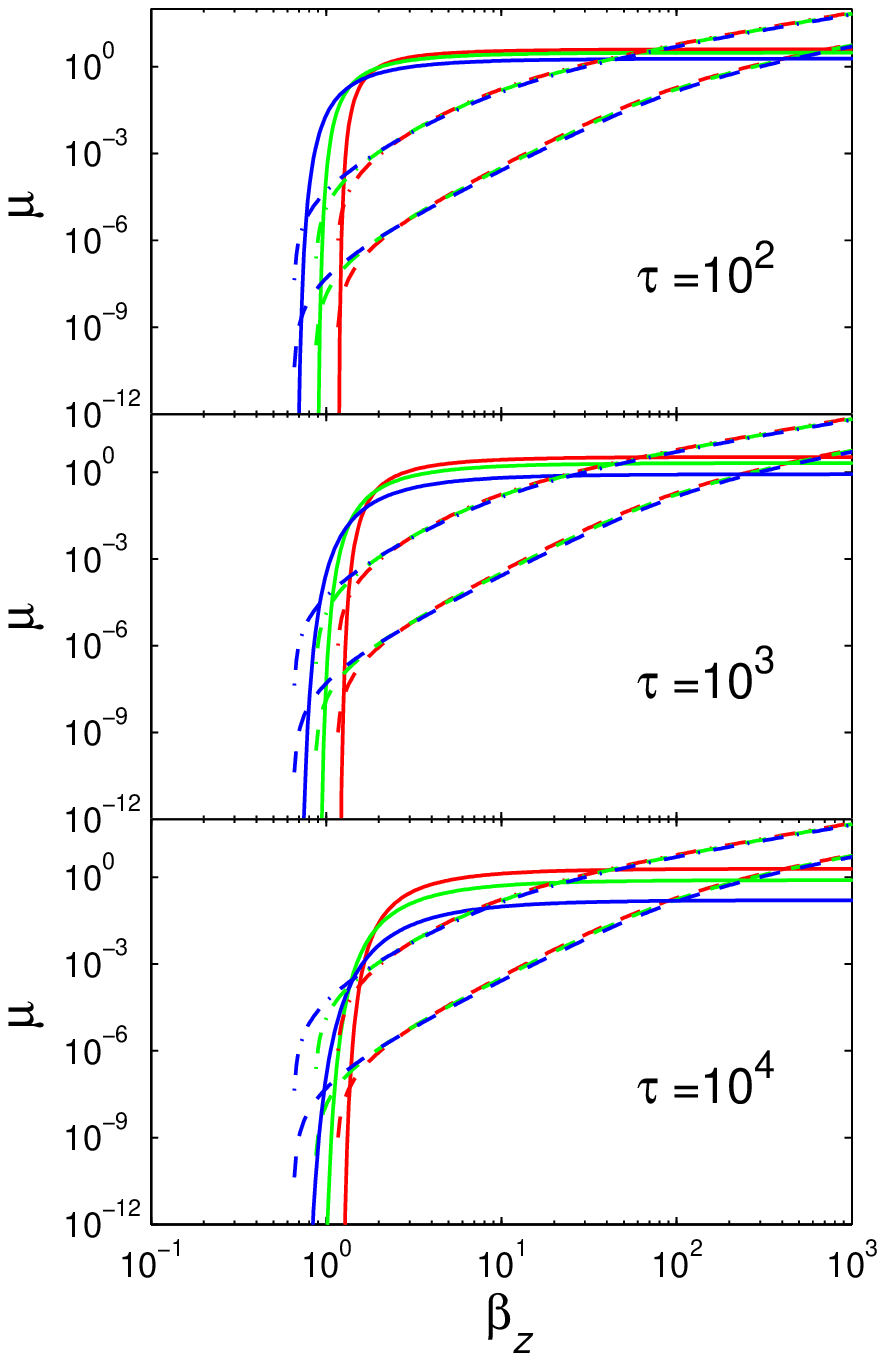}}
\figcaption{Mass loading $\mu$ as a function of $\beta_z$. Solid: as
determined from the wind launching conditions Eq.\
(\ref{mu_launch_2}). Broken: mass loading required for effective
accretion of the field lines (\ref{mu_adv_1g}), for disk viscosity
$\alpha=0.1$ (dashed)) and $\alpha=1$ (dash-dotted). The
intersection points are possible solutions for the stationary wind
driven accretion problem. Field line inclinations are $\kappa_0=1.5$
(red), $\sqrt{3}$ (green), and $2$ (blue). The disk temperature
parameter $\Theta=0.01$, magnetic Prandtl number $\Pm=1$.
\label{beta_mu_th0p1} }\centerline{}
 \vskip 1cm
\figurenum{4}
\centerline{\includegraphics[angle=0,width=7.5cm]{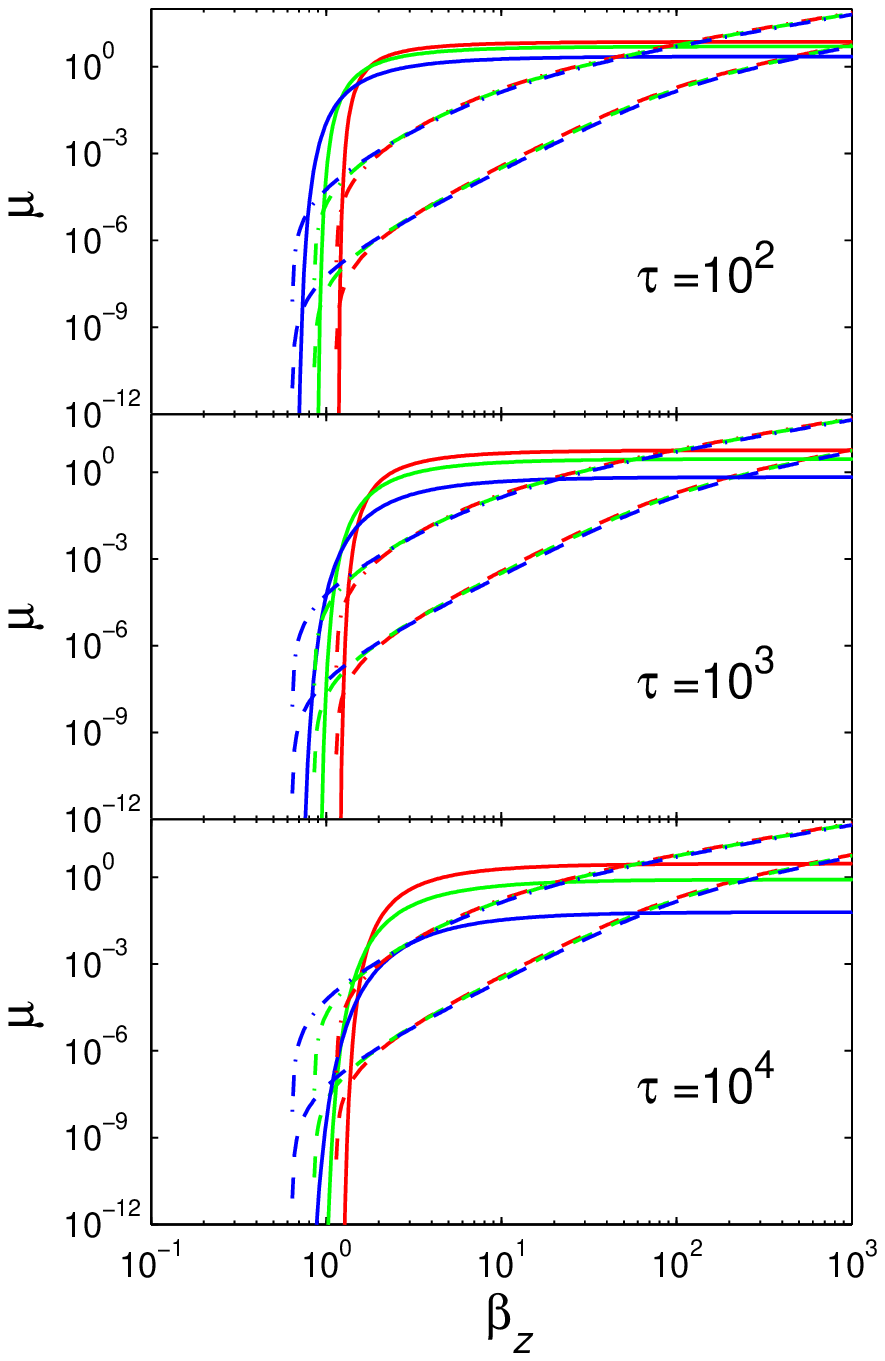}}
\figcaption{As Fig.\ \ref{beta_mu_th0p1} for $\Theta=2.5\times
10^{-4}$. \label{beta_mu_th0p05}  }\centerline{}
 \figurenum{5}
\centerline{\includegraphics[angle=0,width=7.5cm]{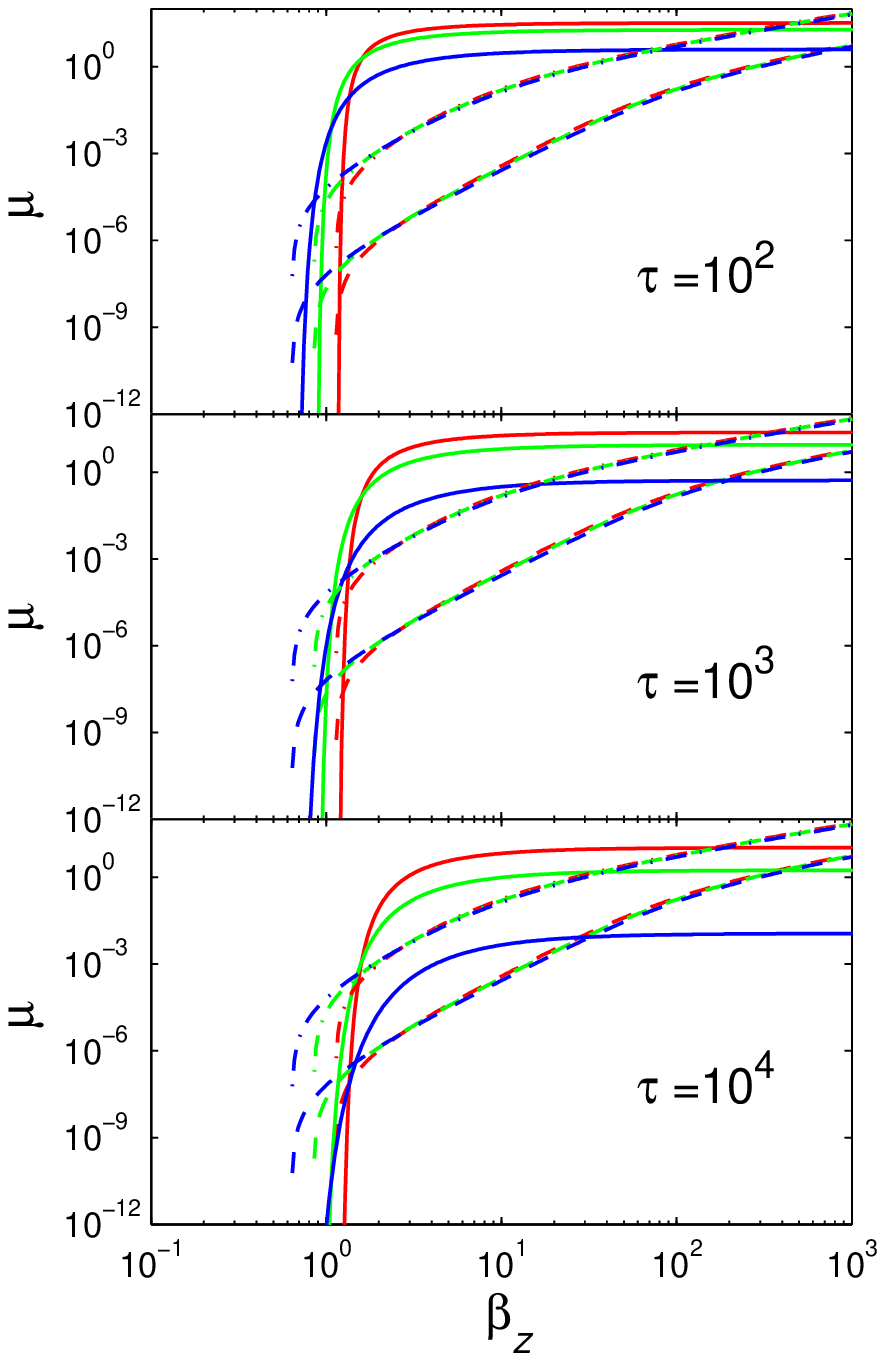}}
\figcaption{As Fig.\ \ref{beta_mu_th0p1}, for $\Theta=10^{-4}$.
\label{beta_mu_th0p01}  }\centerline{} \vskip 1cm
 \figurenum{6}
\centerline{\includegraphics[angle=0,width=7.5cm]{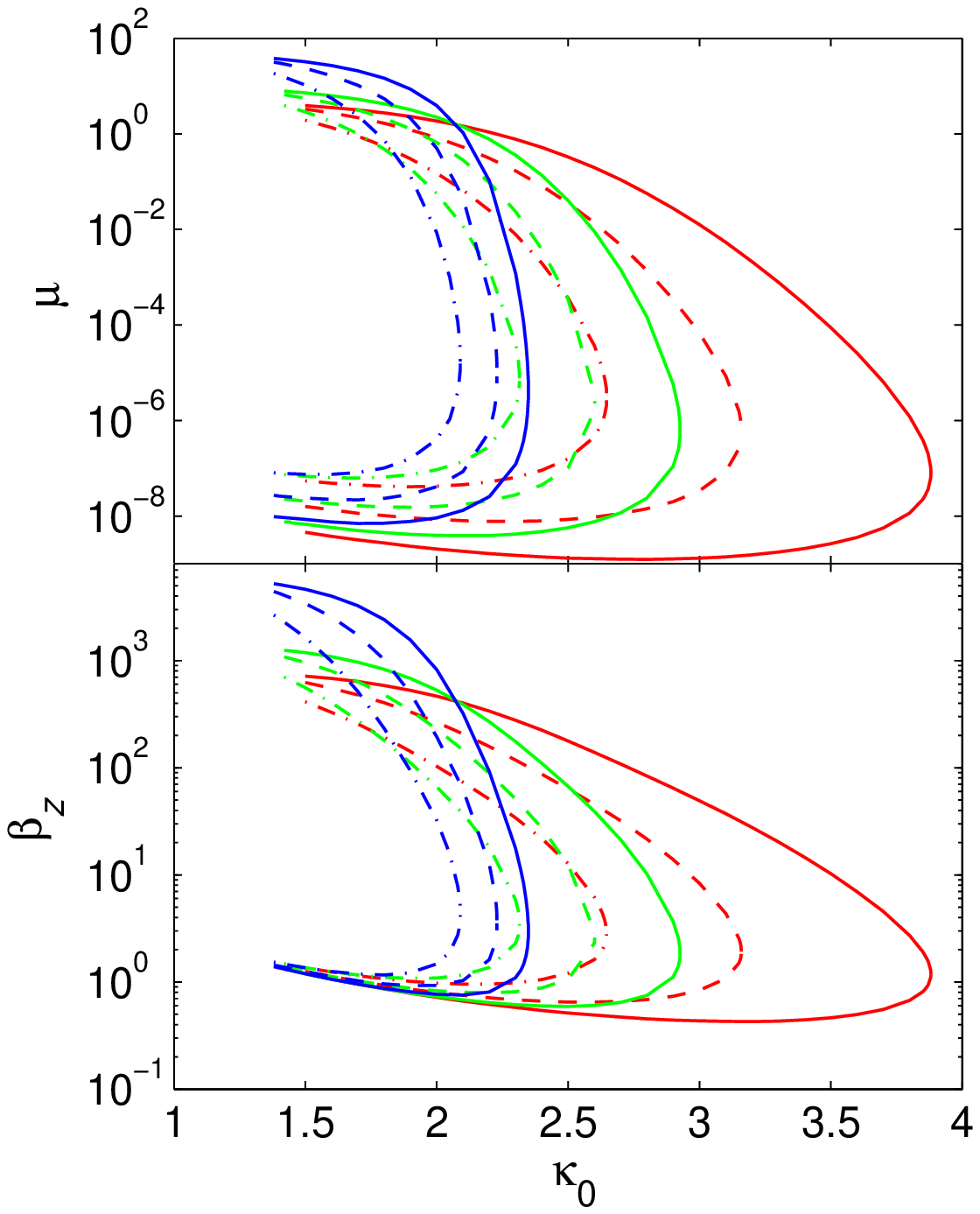}}
\figcaption{Resulting disk-outflow solutions for different disk
temperature parameters: $\Theta=0.01$ (red), $2.5\times 10^{-4}$
(green), and $10^{-4}$ (blue), and different values of the disk
optical depth: $\tau=10^2$ (solid), $10^3$ (dashed), and $10^4$
(dash-dotted). The viscosity parameter $\alpha=0.1$.
 \label{kappa_al0p1} }\centerline{}
\vskip 1cm
 \figurenum{7}
\centerline{\includegraphics[angle=0,width=7.5cm]{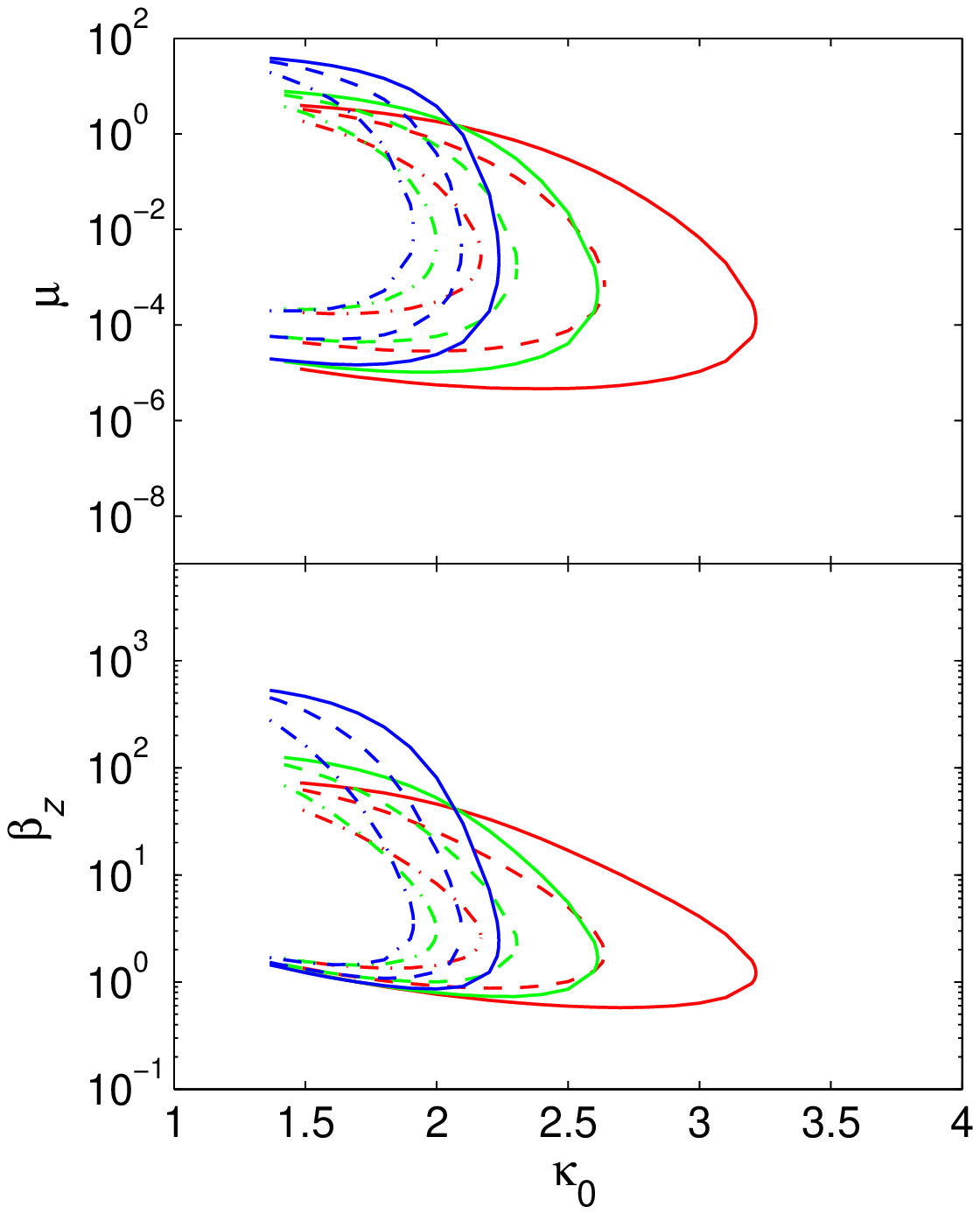}}
\figcaption{As Fig.\ \ref{kappa_al0p1}, for $\alpha=1$.
\label{kappa_al1p0} }\centerline{}
 \vskip 1cm
 \figurenum{8}
\centerline{\includegraphics[angle=0,width=7.5cm]{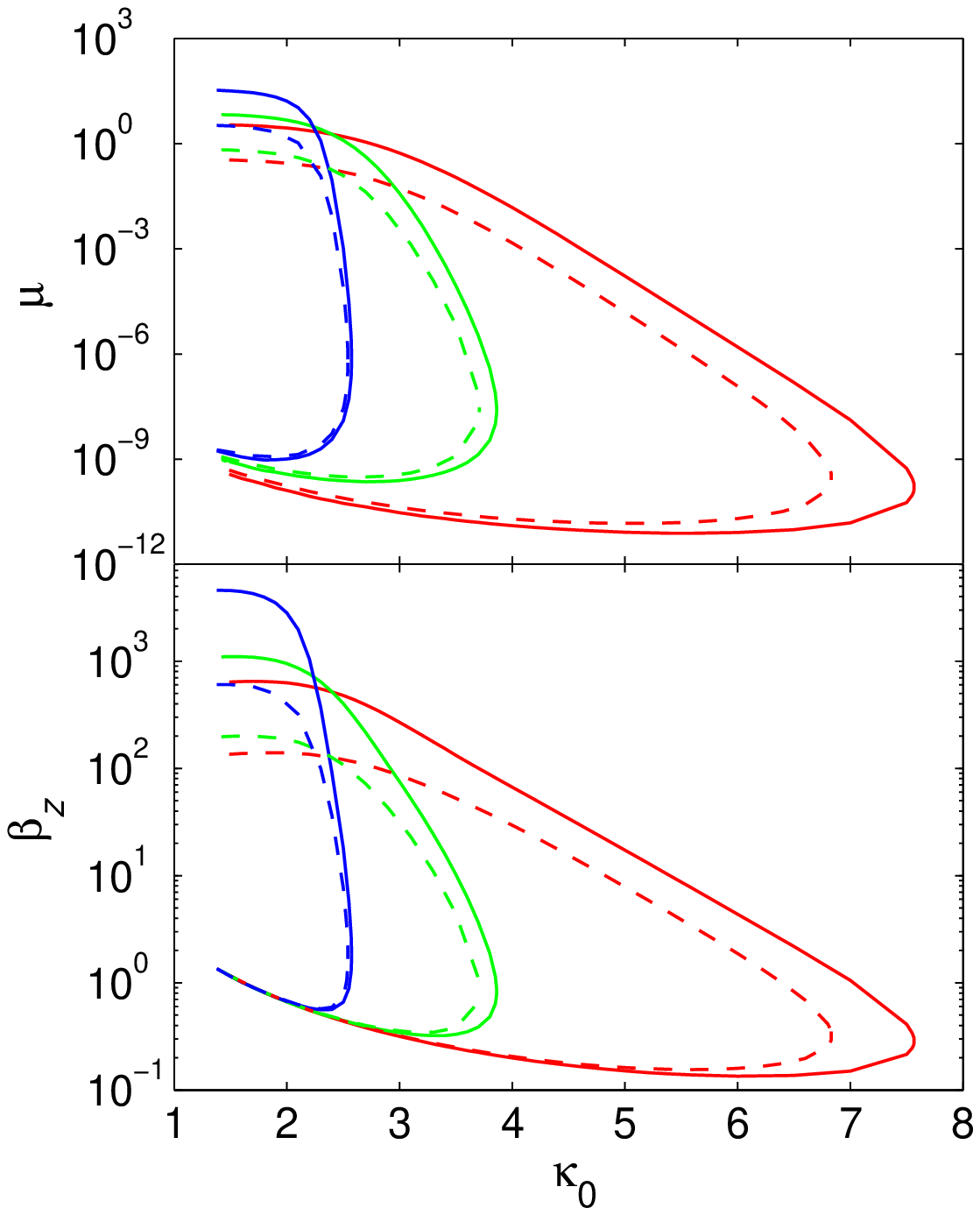}}
\figcaption{The disk-outflow solutions for vertically isothermal
accretion disks. The colored lines represent the solutions derived
with different disk temperature, $\Theta=0.01$ (red), $2.5\times
10^{-4}$ (green), and $10^{-4}$ (blue), respectively. The solutions
derived with different ratios of gas pressure to magnetic pressure
at the disk surface are indicated with different line types,
$\beta_{\rm s}=1$ (solid), and $0.1$ (dashed), respectively.
Viscosity parameter $\alpha=0.1$. \label{kappa_al0p1i}
}\centerline{}
 \vskip 1cm
 \figurenum{9}
\centerline{\includegraphics[angle=0,width=7.5cm]{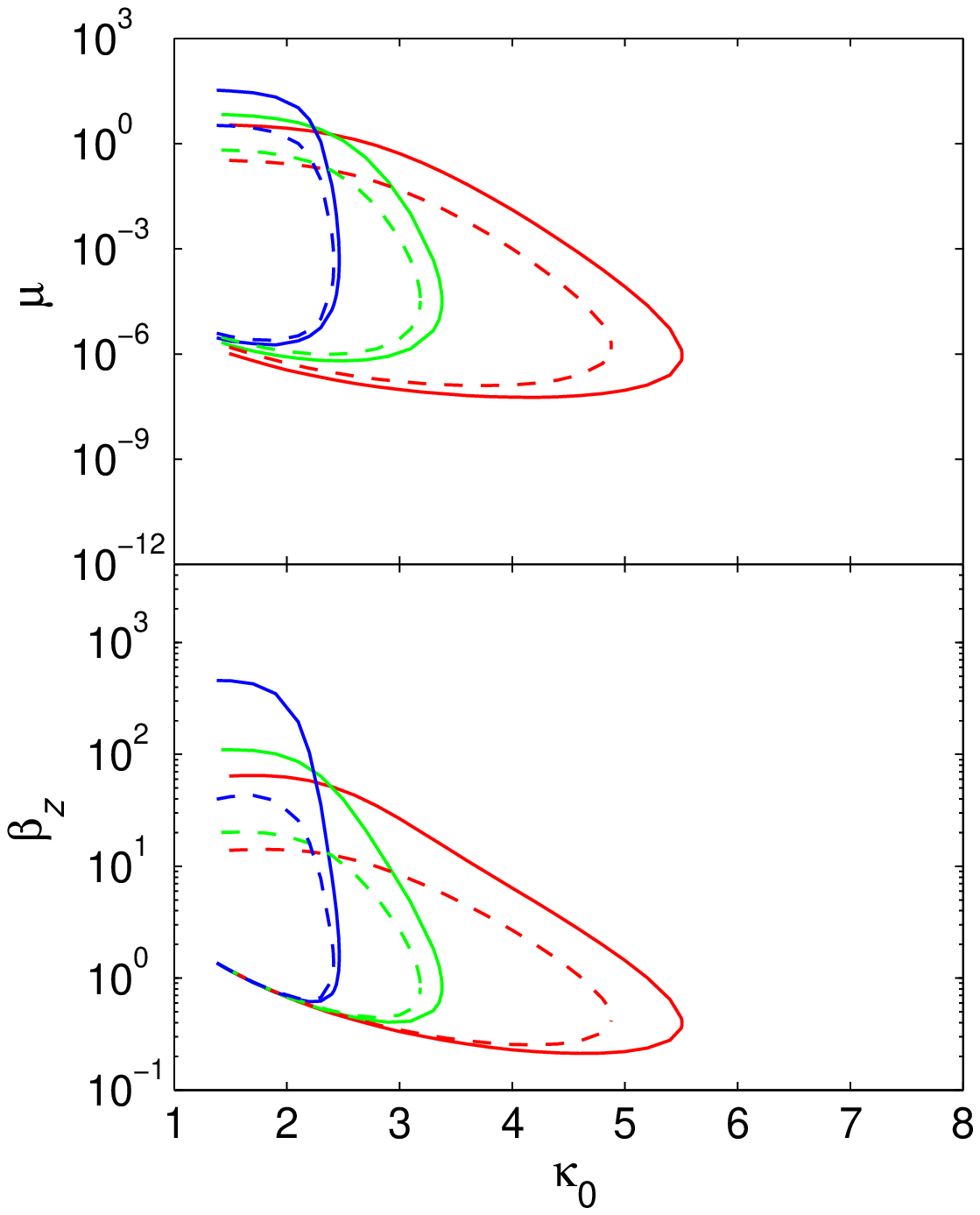}}
\figcaption{The same as Fig.\ \ref{kappa_al0p1i}, for $\alpha=1$.
\label{kappa_al1p0i}  }\centerline{}
 \vskip 1cm
\figurenum{10}
\centerline{\includegraphics[angle=0,width=7.5cm]{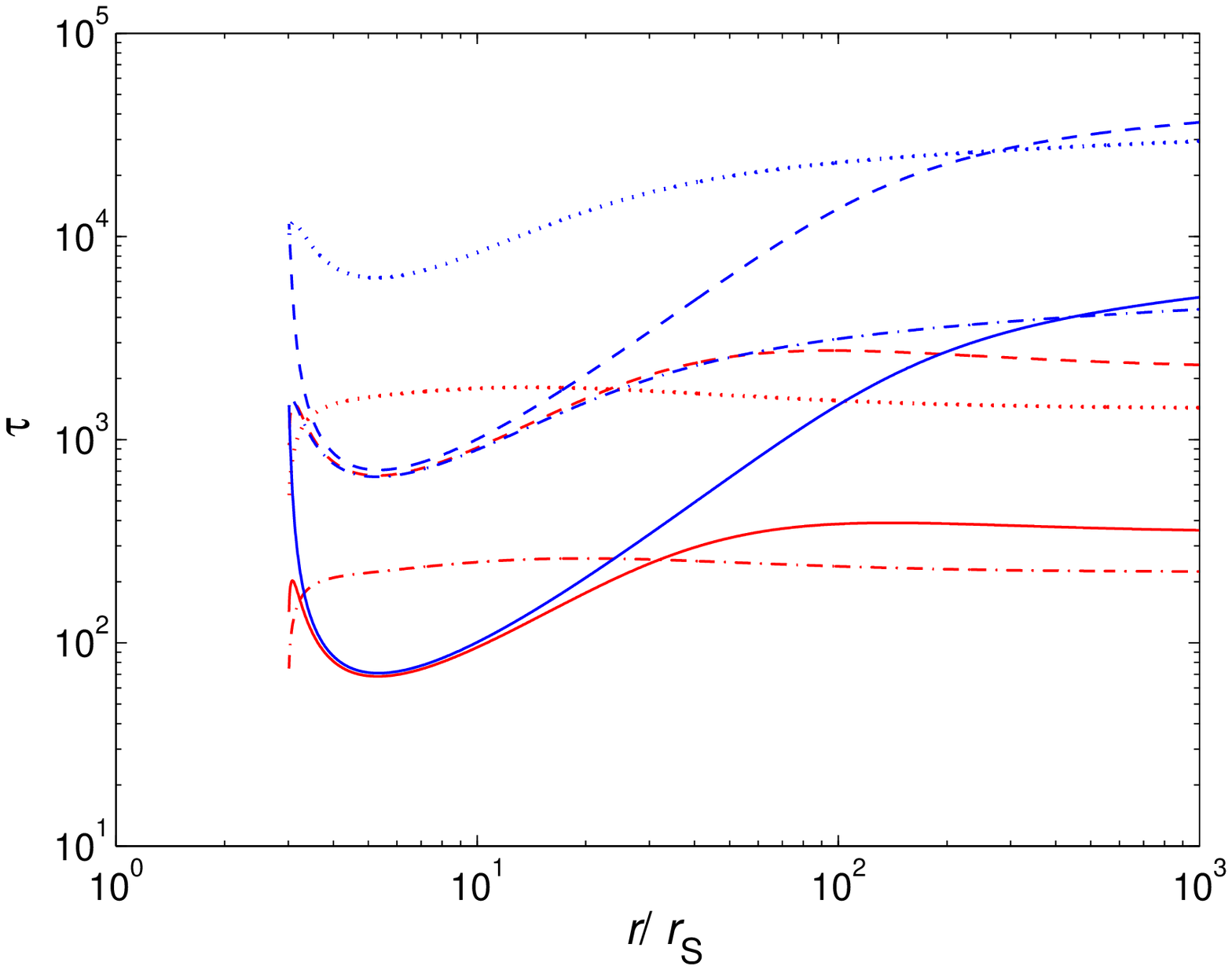}}
\figcaption{The optical depth of the standard thin accretion disks
without magnetic field as functions of radius, where $r_{\rm
S}=2GM/c^2$. The opacity $\kappa_{\rm tot}=\kappa_{\rm
es}+\kappa_{\rm R}$ is adopted, where the Rosseland mean opacity
$\kappa_{\rm R}=5\times 10^{24}\rho T_{\rm
c}^{-7/2}$~cm$^2$~g$^{-1}$. The red lines represent the results
calculated for a black hole with $M=10M_\odot$, while the blue lines
are for a black hole with $M=10^8M_\odot$. The different line types
correspond to the disks with different parameters: solid ($\alpha=1$
and $\dot{m}=0.1$), dashed ($\alpha=0.1$ and $\dot{m}=0.1$),
dash-dotted ($\alpha=1$ and $\dot{m}=0.01$), and dotted lines
($\alpha=0.1$ and $\dot{m}=0.01$). \label{disk_tau} }\centerline{}

The dependence of the solutions on the value of $\alpha$ is shown in
Fig.\  \ref{kappa_al1p0}. The magnetic diffusivity $\eta$ is scaled
with the turbulent viscosity $\nu$, and therefore the diffusion
becomes more important for the cases with a higher value of the
viscosity parameter $\alpha$. In order to compete with the diffusion
of the field, a large radial velocity of the disk is required for
high $\alpha$ cases, which corresponds a high rate of angular
momentum removal by the outflows. We find that the values of the
mass load parameter $\mu$ of the outflow are systematically higher
for those derived with a larger $\alpha$ (compare the results in
Figs.\ \ref{kappa_al0p1} and \ref{kappa_al1p0}).

The accretion disk is vertically compressed by the curved magnetic
field, which sets an upper limit on the field strength (see Eqs.\
\ref{beta_condi_1} and \ref{betaz_condi_1}). The value of
$\beta_{z,\rm min}$ only depends on the field inclination $\kappa_0$
at the disk surface, and $\beta_{z,\rm min}$ decreases with
increasing $\kappa_0$ (see Fig.\ \ref{beta_min}). There is a force
exerted on the disk by the curved field against the gravity of the
central object in the radial direction, which makes the rotation of
the gas in the disk be sub-Keplerian. The rotational velocity of the
disk can be quite low if the field strength is sufficiently strong
(see Eq.\ \ref{omega_2g} and Fig.\ \ref{beta_omega}). The disk is
then magnetically supported against gravity. Such configurations are
likely to be unstable to interchange instabilities, however (Spruit
et al. 1995), which effectively cause the magnetic field to spread
outward and limit the field strength (as observed in the numerical
simulations of Stehle \& Spruit 2001). The requirement that the
magnetic force is less than the gravity in the radial direction
provides an additional constraint on the field strength. We find
that the constraints almost overlap with that constrained by the
force in the vertical direction when $\kappa_0\la 3$, while maximal
field strength becomes lower (a larger $\beta_{\rm min}$) for a disk
with relative high temperature and $\kappa_0\ga 3$ (see Fig.\
\ref{beta_min}).

From Fig.\  \ref{beta_min} we see that the magnetic field can be
very strong, e.g., the magnetic pressure can be more than one order
of magnitude higher than the gas pressure in the disk if the field
inclination $\kappa_0$ is sufficiently large. However, the
magnetically driven outflow will be suppressed if $\kappa_0$ is too
large, because the effective potential barrier becomes extremely
deep in this case. The results show that $\beta\ga 0.5$ is
always satisfied in the disk-outflow solutions (see Figs.\
\ref{kappa_al0p1} and \ref{kappa_al1p0}). Note that this does not
mean the magnetic field cannot be strong, as the disk is
significantly compressed in the vertical direction, which increases
the density of the disk and then the gas pressure for given disk
temperature.

\subsection{Stability}
We have calculated only stationary solutions, but their stability to
time-dependent perturbations can already be guessed at by inspection
of the intersection points in Fig.\ \ref{beta_mu_th0p1}. Near the high-beta
(low field)  solution the mass loading parameter (solid line) is nearly
independent of the field strength assumed (this is because of the
assumed value of the plasma-beta at the base of the flow). If the
field strength were to decrease (to the right of the intersection point),
the mass loss would need to increase in order to maintain a balance
between inward accretion and outward diffusion (dashed line). Since
the mass loading actually does not change much, the angular momentum
loss is insufficient to balance outward diffusion for such a perturbation.
Outward diffusion will then tend to decrease the field strength, providing
a positive feedback to the perturbation. This stationary solution is thus
expected to be unstable.

This mechanism of instability is the same as that identified in the
linear stability analysis of CS02. {The timescale of the instability
is comparable with the dynamical timescale of the disk if the
magnetic torque is large, which becomes significantly small when the
magnetic torque is weak (see CS02 for the detailed results and
discussion). At sufficiently low field strengths or/and
high-$\kappa_0$, where the magnetic torque becomes weak, this
analysis predicted a regime of stability due to magnetic diffusion.
This implies that the growth timescale of such instability
considered in this work should be significantly lower than the
dynamical timescale of the disk. The detailed calculation of the
instability is beyond the scope of this work. }

New is the high-field solution found (leftmost intersection). By the same
line of reasoning as above, this point is expected to be stable, since the
slopes of the dashed and solid curves are reversed here. It is,
however, somewhat outside the assumptions made, since the MRI
turbulence that was assumed for the magnetic diffusion is probably
suppressed  at these field strengths. Instead, instability of the strong
field itself is likely to cause its outward diffusion (as in the simulations of
Stehle \& Spruit 2001 and Igumenschev et al.\ 2003). To the (uncertain) extent
that this process can be parametrized in terms of $\cs^2/\Omega$, the
present analysis would still apply. We speculate that the stability of this
point is actually significant, and that it is relevant for the experimentally
observed stability of the strong central flux bundles in numerical simulations
of  accretion onto black holes.

\section{Conclusions}
We have considered the possibility that the angular momentum of an
accretion disk is removed predominantly by outflows driven
by the accreted field (Bisnovatyi-Kogan \& Ruzmaikin 1974,
Blandford 1976). An obstacle to this proposal has been the
realization (van Ballegooijen 1989, \citet{1994MNRAS.267..235L})
that outward diffusions of the accreted field is fast compared to the
inward accretion velocity in a geometrically thin accretion disk if the
value of the Prandtl number $\Pm$ is around unity. Revisiting this
problem, we find that even moderately weak fields can in fact cause
sufficient  angular momentum loss via a magnetic wind to balance
outward diffusion. The estimate in Eq.\ (\ref{mu_adv_1}) shows that,
at  $\Pm=1$, a field with a magnetic pressure as low as $\sim 10^{-3}$
of the gas pressure $p_{\rm c}$ at the disk midplane has a chance
of facilitating its own accretion by driving a moderately strong
magnetic outflow. This is due  more or less to compounding numerical
factors of order unity.  In particular when  MRI turbulence produces the
relatively low effective viscosity $\alpha\sim 0.01$ that is seen in several
numerical simulations.

Using a simplified model for the vertical disk  structure, we have
studied the conditions for existence of such stationary equilibria
between wind-induced advection and outward turbulent diffusion in
more quantitative detail. Two equilibrium points are found, one at
low field strengths corresponding to a plasma-beta at the midplane
of order several hundred, and one for strong accreted fields,
$\beta\sim 1$. We surmise that the first is relevant for the
accretion of weak, possibly external, fields through the outer parts
of the disk, while the latter one could explain the tendency,
observed in full 3D numerical simulations, of strong flux bundles at
the centers of disk to stay confined in spite of strong MRI
turbulence surrounding them (e.g.\ Beckwith et al. 2009). {These
authors also identify the mechanism responsible for maintenance of
the bundle against the outward diffusion that one might expect from
the turbulence surrounding it. Unlike the present model, this
mechanism does not depend on the presence of a wind.}

\acknowledgments HS thanks the Shanghai Astronomical Observatory for
their generous hospitality during the work on the project reported
here. This work is supported by the National Basic Research Program
of China (grant 2009CB824800), the NSFC (grants 11173043, 11121062
and 11233006), and the CAS/SAFEA International Partnership Program
for Creative Research Teams (KJCX2-YW-T23).

{}

\end{document}